\documentclass[conference,compsoc]{IEEEtran}

\ifCLASSOPTIONcompsoc
  \usepackage[nocompress]{cite}
\else
  \usepackage{cite}
\fi
\usepackage[pdftex]{graphicx}
\usepackage{color}
\usepackage{braket}
\usepackage{amsmath,amssymb,bbm}
\usepackage{booktabs,array}
\usepackage{epstopdf}
\usepackage{placeins}

\graphicspath{{figures/plots/}{figures/}{figures/gates/}}

 \usepackage[table,xcdraw]{xcolor}

\begin{document}

\title{High Performance Emulation of Quantum Circuits}

\author{\IEEEauthorblockN{Thomas H\"aner\IEEEauthorrefmark{1},
Damian S. Steiger\IEEEauthorrefmark{1},
Mikhail Smelyanskiy\IEEEauthorrefmark{2}, 
and
Matthias Troyer\IEEEauthorrefmark{1}\IEEEauthorrefmark{3}\IEEEauthorrefmark{4}}
\IEEEauthorblockA{\IEEEauthorrefmark{1}Institute for Theoretical Physics, 
ETH Zurich,
8093 Zurich, Switzerland}
\IEEEauthorblockA{\IEEEauthorrefmark{2} Parallel Computing Lab, 
 Intel Corporation}
\IEEEauthorblockA{\IEEEauthorrefmark{3} Quantum Architectures and Computation Group, Microsoft Research, Redmond, WA (USA)}
 \IEEEauthorblockA{\IEEEauthorrefmark{4} Corresponding Author: 
 troyer@phys.ethz.ch}}

\maketitle

\begin{abstract}
As quantum computers of non-trivial size become available in the near future, it is imperative to develop tools to emulate small quantum computers. This allows for validation and debugging of algorithms as well as exploring hardware-software co-design to guide the development of quantum hardware and architectures. The simulation of quantum computers entails multiplications of sparse matrices with very large dense vectors of dimension $\mathbf{2^n}$, where $\mathbf n$ denotes the number of qubits, making this a memory-bound and network bandwidth-limited application. We introduce the concept of a quantum computer \textit{emulator} as a component of a software framework for quantum computing, enabling a significant performance advantage over simulators by emulating quantum algorithms at a high level rather than simulating individual gate operations. We describe various optimization approaches and present benchmarking results, establishing the superiority of quantum computer emulators in terms of performance.
\end{abstract}

\IEEEpeerreviewmaketitle

\section{Introduction}

Sustaining the pace of Moore's law~\cite{Moore:2000:CMC:333067.333074} has become increasingly difficult over recent years. Despite the significant amount of innovation in the last decade, which helps to meet the power and performance requirements of exa-scale systems, it is likely that an alternative approach will enable the steady growth of computing power beyond exa-scale.

One such alternative may be quantum computers, which offer the potential of exponential  speedup for certain types of calculations. In recent years, many quantum algorithms with substantial speedup over the best known classical algorithms have been developed \cite{jordan2011quantum}, with applications ranging from factoring large numbers~\cite{shor94} to quantum chemistry~\cite{aspuru2005simulated} and materials science~\cite{bauer2015hybrid}. Due to the fact that the majority of high-performance computing time is spent to solve problems in chemistry and materials science, this approach is especially lucrative and worth of further investigation, as it could result in sustained performance improvements of these applications.

While large-scale quantum computers are not yet available, their performance can be inferred using quantum compilation frameworks and estimates of potential hardware specifications. However, without testing and debugging quantum programs on small scale problems, their correctness cannot be taken for granted. Simulators and emulators of small quantum computers are essential to address this need. Specifically, they increase the productivity in the field of quantum algorithm development by facilitating the testing, debugging, and exploration of new algorithms before large-scale quantum computers become available. Furthermore, they allow for hardware-software co-design and guide the development of quantum hardware architectures.

The state of a quantum computer with $n$ qubits can be represented by a state vector of $2^n$ complex amplitudes. Operations on qubits, called gates, correspond to multiplications of this state vector with (typically) sparse unitary matrices of dimension $2^n\times2^n$. Due to the exponential overhead of simulating a quantum system on a classical computer, it is  crucial to make use of parallelism and all possible optimizations in order to reduce runtime and enable the simulation of as many qubits as possible.
\\\\
\textbf{Our contribution.} In this paper, we introduce the concept of a quantum computer \textit{emulator}, which extends the notion of a simulator to the case where a comprehensive compilation framework for quantum programs is available \cite{haener2016}, enabling an entirely new class of optimizations useful for the simulation of quantum computers. An emulator makes use of the availability of an abstract, high-level quantum code by directly employing classical emulation for quantum subroutines at the level of their mathematical description instead of compiling them into elementary gates prior to applying them using a series of sparse matrix vector multiplications. As a consequence, the overall runtime of simulating the execution of quantum programs can be drastically reduced, arriving at an unprecedented level of performance.

We present various examples of such optimizations, accompanied by performance measurements showing the merits of quantum computer emulation. Furthermore, in order to arrive at heuristics for cases where multiple classical shortcuts exist, an analysis of cross-over points is carried out. Finally, to demonstrate that our simulator, against which we achieve a speedup using our new quantum emulator, is state-of-the-art, we benchmark it against other existing simulators on a subset of quantum circuits, and show its superior performance.
\\\\
\textbf{Related work.}
Emulation is a widely recognized tool commonly used in many areas of computer science. A fitting example is the $\text{Intel}^{\textregistered}$  Software Development Emulator tool~\cite{Luk:2005:PBC:1065010.1065034}, which allows fast emulation of upcoming or experimental hardware features, such as new SIMD or transactional memory extensions, before they become available. We extend this approach to the domain of quantum computers.
\\\\
\textbf{Outline.}
This paper is organized as follows: In Section~\ref{sec:qcsim} we give an introduction to quantum computer simulation, followed by a description of the tricks an emulator can use in order to gain speedups over simulators in Section~\ref{sec:qcemul}. In Section~\ref{sec:results}, we present timings supporting this claim and compare the performance of our simulator to other state-of-the-art simulators in order to assert that the speedup stems from emulation and not from a badly implemented simulator. Finally, we summarize our findings in Section~\ref{sec:summary}.

\section{Quantum Computer Simulation}\label{sec:qcsim}

Information on a quantum computer is stored in the quantum generalization of bits, called qubits, and computations are performed by applying quantum gates and measurement operations to these qubits, similar to how classical gates are applied to classical bits.

In this paper, we discuss the simulation of an ideal quantum computer on a logical level, ignoring all noise which would occur in an actual physical device. In this case, a quantum computer with $n$ qubits can be simulated by representing the  quantum state (the so-called ``wave function'') of the qubits as a vector of size $2^n$ with complex entries $\alpha_i$ ($i\in\{0,1\}^n$) satisfying the normalization condition $\sum_i \left|\alpha_i\right|^2 = 1$. A general state of the $n$ qubits is thus represented by a vector of the form
\begin{equation}
	\alpha=\left(\begin{matrix}\alpha_0 & \alpha_1 & \dots & \alpha_{2^n-1}\end{matrix}\right)^T.
\end{equation}

Information stored in qubits is retrieved using measurements, which convert qubits into classical bits. While the quantum state is, in general, a superposition of all classical states $i$, a measurement of all qubits ``collapses'' this state to a single classical $n$-bit configuration $i$, where $i$ is chosen randomly with probability $\left| \alpha_i \right|^2$.

In this representation, the application of quantum gates to qubits results in a modification of the $2^n$ complex entries of the $n$-qubit state vector. Fundamental physical principles dictate that every quantum gate is required to be unitary, which means that it can be represented by a unitary matrix of dimension $2^n\times2^n$. The application of a quantum gate then corresponds to a multiplication of the $n$-qubit state vector with the unitary gate matrix, which is sparse for gates acting on a small number of qubits.

Most experimental implementation of quantum computers are only capable of performing operations on one or two qubits. This is not a limitation as there exist sets of one- and two-qubit quantum gates which are universal for quantum computation. Therefore, most quantum algorithms are decomposed into one- and two-qubit gates, which can each be represented by  $2^n\times2^n$ sparse unitary matrices. They are often expressed in an even more compact form. One qubit gate operations, for example, can be represented by $2\times2$ matrices acting on the state of a single qubit. The $2^n\times2^n$ size unitary matrix acting on the $2^n$ complex entries of the state vector can then be constructed from these small matrices. A few of the most frequently used gate matrices and symbols are shown in Table~\ref{tbl:gates}. The NOT operation, for example, is given by
\begin{equation}
\left(\begin{matrix}0 & 1\\1 & 0\end{matrix}\right)\;.
\end{equation}
As an example on how to construct the $2^n\times2^n$ matrix from this, consider applying the quantum NOT gate to qubit $i$. The matrix corresponding to this operation is equal to the Kronecker product of a series of matrices $m_j$ $(0\leq j<n)$ of size $2\times2$, where $m_j$ is equal to the identity matrix except for $j=i$, where it is the NOT gate matrix instead. For a quantum computer with $n=2$ qubits, the matrix corresponding to a NOT operation on qubit $j=0$ is therefore given by
\begin{equation}
X\otimes\mathbbm 1_2=\left(\begin{matrix}0 & 1\\1 & 0\end{matrix}\right)\otimes\left(\begin{matrix}1 & 0\\0 & 1\end{matrix}\right)=\left(\begin{matrix}
0 & 0 & 1 & 0\\
0 & 0 & 0 & 1\\
1 & 0 & 0 & 0\\
0 & 1 & 0 & 0
\end{matrix}\right)\;.
\end{equation}
From this construction it becomes obvious that one- and two-qubit gates are very sparse $2^n\times2^n$ matrices with compile-time fixed entries. Hence a simulator can apply various low-level optimization strategies compared to generic sparse matrix vector multiplication, including optimizing away multiplications by ones and zeros and reducing communication in distributed implementations. While this can improve the performance of the simulation by a fair amount, it is still no match for an emulator, as we will show in the next sections. 

\begin{table}
	\centering
	\renewcommand{\arraystretch}{1}
	\begin{tabular}{>{\centering}m{2cm} >{\centering}m{2cm} >{\centering\arraybackslash}m{2cm}}
	\toprule
	Gate & Matrix & Symbol\\\hline
 NOT or X gate& \(\displaystyle\left(\begin{matrix}0 & 1\\1 & 0\end{matrix}\right)\)&\includegraphics[width=2cm]{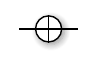}\\
 Y gate & \(\displaystyle\left(\begin{matrix}
 			0 & -i\\
 			i & 0
 			\end{matrix}\right)\)&\includegraphics[width=2cm]{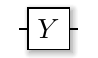}\\
	Z gate & \(\displaystyle\left(\begin{matrix}
			1 & 0\\
			0 &-1
			\end{matrix}\right)\)&\includegraphics[width=2cm]{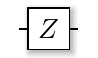}\\
	Hadamard gate & \(\displaystyle\frac 1{\sqrt2}\left(\begin{matrix}
			1 & 1\\
			1 &-1
			\end{matrix}\right)\)&\includegraphics[width=2cm]{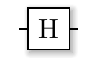}\\
	S gate& \(\displaystyle\left(\begin{matrix}
			1 & 0\\
			0 & i
			\end{matrix}\right)\)&\includegraphics[width=2cm]{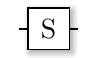}\\
	T gate & \(\displaystyle\left(\begin{matrix}
			1 & 0\\
			0 & e^{i\pi/4}
			\end{matrix}\right)\)&\includegraphics[width=2cm]{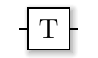}\\
	Rotation-Z gate & \(\displaystyle\left(\begin{matrix}
			e^{-i\theta/2} & 0\\
			0 & e^{i\theta/2}
			\end{matrix}\right)\)&{\centering\includegraphics[width=2cm]{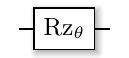}}\\
	Controlled NOT (CNOT) & \(\displaystyle\left(\begin{matrix}
				1 & 0 & 0 & 0\\
				0 & 1 & 0 & 0\\
				0 & 0 & 0 & 1\\
				0 & 0 & 1 & 0
				\end{matrix}\right)\)&\includegraphics[width=2cm]{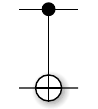}\\
	Conditional phase-shift (CR) & \(\displaystyle\left(\begin{matrix}
				1 & 0 & 0 & 0\\
				0 & 1 & 0 & 0\\
				0 & 0 & 1 & 0\\
				0 & 0 & 0 & e^{i\theta}
				\end{matrix}\right)\)&\includegraphics[width=2cm]{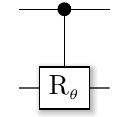}\\\bottomrule\\
	\end{tabular}
	\caption{Standard quantum gates with their corresponding matrices and symbols.}
	\label{tbl:gates}
\end{table}

\section{Quantum Computer Emulation}\label{sec:qcemul}

Simulation and emulation of quantum computers are inherently different concepts. As a \textit{simulation} of a quantum computer we understand the exact calculation of the effects of every single gate. This directly mimics the operations that a quantum  computer performs and the simulation can also include effects of classical and quantum noise as well as calibration or control errors. 

Quantum computer \textit{emulation}, on the other hand, is only required to return the same result as a perfect quantum computation would. Instead of compiling an algorithm down to elementary gates for specific quantum hardware, certain high-level subroutines can be replaced by calls to faster classical shortcuts to be executed by the emulator. Depending on the level of abstraction at which the emulation is carried out, there is a large potential for optimizations and substantial speedup \cite{haener2016}. 

To illustrate this point, consider performing classical functions on a quantum computer which is needed in order to apply classical functionality to a superposition of inputs. The most famous application of this is Shor's algorithm \cite{shor94}.
In order to satisfy the reversibility constraint of quantum mechanics, these functions need to be implemented reversibly, which leads to a large overhead in the number of quantum gates compared to a non-reversible classical computation. This is due to the fact that temporary variables need to be reset by employing a so called uncomputation step \cite{bennett73}.

A straight-forward approach to translating a classical function to a reversible quantum circuit is to replace all NAND gates by the reversible Toffoli gate (also called CCNOT), which requires an additional bit for each NAND to store the result. After completion of the circuit, the result can be copied using CNOT gates prior to clearing all (temporary) work bits by running the entire circuit in reverse \cite{bennett73}. This can also be run on a quantum computer using a quantum version of the Toffoli gate (which can be composed from single-qubit gates and CNOT gates).
This transformation causes a doubling of gates and an overhead of one additional qubit for each original NAND gate. There are more sophisticated approaches \cite{parent15} which reduce the number of work qubits by uncomputing intermediate results early. Yet, those intermediate results have to be recomputed\footnote{As the uncomputation of an uncomputate step is recomputing the original result.} during the uncomputation step which follows after completion of the circuit, resulting in an increase of gate operations. This is bad news for a simulator, since both approaches cause a significant increase in runtime. Hence, the simulation of a classical function on a quantum computer is a very costly endeavor.

An emulator, on the other hand, does not need to compile the classical function down to reversible gates, nor does it have to simulate the additional work qubits that may be needed during function execution. Instead, the emulator can just evaluate the classical function directly for each of its arguments, thereby saving huge amounts of computational power. Below we will discuss four examples where emulation may gain a substantial performance advantage over simulation.

\subsection{Arithmetic Operations and Mathematical Functions}
The most straight-forward example is the execution of arithmetic operations and mathematical functions on a quantum computer. Instead of simulating the vast number of Toffoli gates required to implement {\it e.g.} a multiplication or a trigonometric function reversibly, one can perform the classical multiplication or trigonometric function directly for each computational basis state using the hardware implementation available on classical computers.

We consider the multiplication and division of two numbers $a$ and $b$ into a new register $c$ as examples. Specifically, we implement the mapping for multiplication
\begin{align*}
	(a,b,c=0)&=
	(a_1,...,a_N,b_1,...,b_N,0,...,0)\\&\mapsto(a,b,ab)\;,
\end{align*}
and for division (with remainder $r$), 
\begin{align*}
	(a,b,c=0)&=
	(a_1,...,a_N,b_1,...,b_N,0,...,0)\\&\mapsto
	(r,b,a/b)\;,
\end{align*}
where the $N$-qubit input registers $a$ and $b$ may be in an arbitrary superposition, allowing this computation to be carried out on all (exponentially many) possible input states in parallel on a quantum computer.

On a simulator, the $3N$-qubit wavefunction is stored as a vector of $2^{3N}$ complex numbers with indices $i\in\{0,1\}^{3N}$, which can be written as $i=a_1,...,a_N,b_1,...,b_N,0,...,0$ in binary notation, where $x_k$ denotes the $k$-th bit of $x$. The action of a multiplication corresponds to a permutation of the state vector, mapping the complex value at location $i$ to the index $j=a_1,...,a_N,b_1,...,b_N,(ab)_1,...,(ab)_N$.
In order to achieve this transformation, a simulator would apply the corresponding Toffoli network. An emulator, on the other hand, can simply perform the described mapping directly. The simulation and emulation of a division can be carried out analogously.

For benchmarking the simulation, we implement these operations using the adder of Ref.~\cite{cuccaro2004new} combined with a repeated-addition-and-shift and a repeated-subtraction-and-shift approach for multiplication and division, respectively. The runtimes of emulation and simulation can be found in section~\ref{sec:results}.

\subsection{Quantum Fourier Transform}
The quantum Fourier transform (QFT) \cite{nielsen2010} is a common quantum subroutine that is used in many quantum algorithms due to its ability to detect periods and patterns. At a formal mathematical level, the QFT performs a Fourier transform on the state vector $\alpha$ of $n$ qubits, where each entry $\alpha_l$ ($0\le l < 2^n$) gets transformed as
\begin{equation}
	\alpha_l\mapsto\frac 1{2^{n/2}}\sum_{k=0}^{2^n-1} \alpha_k \exp\left(2\pi i\frac{k l}{2^n}\right)\;.
\end{equation}
On a quantum computer, the QFT can be implemented by a sequence of $\mathcal{O}(n^2) $ Hadamard and conditional phase shift gates.
Simulating this circuit is expensive as each gates acts on the state vector of size $2^n$. An emulator, on the other hand, can just directly perform a Fast Fourier Transform (FFT) on the state vector using optimized classical libraries. 

We consider performance models of both algorithms on a distributed system in order to understand the advantage of FFT over QFT. Let $T_{\text{FFT}}(n)$ and $T_{\text{QFT}}(n)$ denote the execution times of node-local FFT and QFT, respectively, where $N=2^n$ is the size of the input vector. It is well known that a all industry high-performance implementation of a distributed 1D FFT require three all-to-all broadcasts (see ~\cite{Bailey:1990:FEH:81777.81781} and ~\cite{FFTW05}, for example), due to three transposition steps. Hence, its performance can be modeled as follows:

\begin{equation}
 T_\text{FFT}(n) = \frac{5  N  n}{ \text{Eff}_\text{FFT} \times \text{FLOPS}_\text{peak}} + 3  \frac{16N}{\text{B}_\text{net}}\;,
 \label{eq:tfft}
\end{equation}
where $\text{Eff}_\text{FFT}$ denotes the efficiency of the FFT, which typically ranges between $10\%-20\%$ on current architectures and $B_{net}$ is the aggregate injection bandwidth of the distributed system in question.

The direct simulation of a QFT, on the other hand, requires the application of $n$ single-qubit Hadamard gates and $n(n-1)/2$ controlled phase shift operations, which thus dominate the execution time. Furthermore, phase shifts are diagonal matrices with the first diagonal entry of 1 (see Table~\ref{tbl:gates}). Thus we implement this controlled operation by reading and writing only a quarter of the double complex state vector. This results in approximately 
$
	\frac{n^2}{2}\text{ [operations]} \cdot 2 \text{ [accesses (read/write)]} \cdot 16 \text{ [bytes/entry]} \cdot\frac{N}{4} \text{ [accessed entries]}=4\cdot N \cdot n^2
$
bytes of data being accessed. This results in compute time $\frac{4 N  n^2}{\text{B}_\text{mem}}$, where $B_{mem}$ is the aggregate memory bandwidth of the system.

Only the application of Hadamard gates to high-order qubits $k$, where $k \ge (n-\log_2(P))$ and $P$ is number of nodes, requires communication. This occurs $\log_2(P)$ times during the application of the entire QFT. 
Thus QFT performance is:
\begin{equation}
T_\text{QFT}(n) = \frac{4 N  n^2}{\text{B}_\text{mem}} + \log_2(P) \frac{16 N}{\text{B}_\text{net}},
\label{eq:tqft}
\end{equation}

Note that QFT scales worse than FFT in both computational and communication complexity under ideal assumptions. Therefore, we expect the performance advantage of FFT to grow with the number of qubits.

\subsection{Quantum Phase Estimation}
\label{subsec:qpe}

Quantum phase estimation (QPE) \cite{nielsen2010} is another subroutine that is used in many quantum algorithms, such as Shor's algorithm for factoring \cite{shor94}.
Given a circuit of a unitary operator $U$ acting on $n$ qubits and an eigenvector $\vec{u}$ stored in an $n$-qubit register, the QPE algorithm calculates the corresponding eigenvalue $e^{i\theta}$.\footnote{All eigenvalues of a unitary matrix can be written in the form of $e^{i\theta}$. QPE calculates the angle $\theta$.} In a wave function simulator picture, the operator $U$ is described by a unitary $2^n\times 2^n$ matrix. While there are many versions of the QPE algorithm, they all are based on repeatedly applying the controlled operator $U$.\footnote{Since the cost of simulating the application of the controlled version of $U$ is essentially the same as simulating $U$ itself, we ignore this detail in the further analysis.} Specifically, the operators
\begin{align}
\label{eq:powers_of_u}
	U^1, U^2, U^4, U^8,\dots,U^{2^{b-1}}
\end{align}
need to be applied in order to arrive at a $b$-bit estimate of the eigenvalue angle $\theta$.  In addition, at least one work qubit and an inverse QFT are required when implementing the QPE as done in Ref.~\cite{beauregard02}. In the following, we assume that $U$ is implemented on a quantum computer through a sequence of $G$ gates, i.e.,
\begin{align*}
U=\prod_{i=1}^G U_i\;,
\end{align*}
where $U_i$ is a single or two-qubit gate.

A simulator implements $U_i$ through multiplications of sparse  $2^{n}\times2^{n}$ matrices with the wave function. Applying powers of $U$ corresponds to repeatedly applying the sequence of $G$ gates which each has complexity $\mathcal{O}(2^{n})$. From equation~\ref{eq:powers_of_u}, it follows that we need to apply $U$ exactly $2^b-1$ times. Hence, the runtime complexity of QPE without accounting for the inverse QFT is $\mathcal{O}(G2^{n+b})$ for a quantum computer simulator using an algorithm with the minimal number of one ancilla qubit as done in \cite{beauregard02}.
Coherent phase estimation algorithms~\cite{nielsen2010} that use $b$ ancillas to optimize runtime will incur an additional factor $\mathcal{O}(2^b)$ in simulation effort.

An emulator can take a shortcut by first building a (dense) matrix representation of the unitary operator $U$ and then using repeated squaring to calculate $U^{2^i}$ iteratively for $i=0,1,...,b-1$.  Building the matrix representation of $U$ requires $\mathcal{O}(G2^{2n})$ effort. Using standard matrix-matrix multiplication, repeated squaring can be performed in time $\mathcal{O}(2^{3n} b)$. Using Strassen's algorithm, the complexity can be reduced to $\mathcal{O}(2^{2.8n} b)$. Since $G$ is typically polynomial in $n$, $G2^{n}$  is sub-dominant. There is an advantage in the asymptotic scaling when switching from quantum simulation to emulation if $b \ge 2n$, or $b>(\log_27-1)n \approx 1.8n$ when using Strassen.

Alternatively, a dense matrix eigensolver can be employed to directly classically compute the eigenvalues of $U$ with effort $\mathcal{O}(G2^{2n}+2^{3n})$ for approaches based on Hessenberg reduction  \cite{golub1979hessenberg}, which again will have a scaling advantage compared to simulation for $b>2n$. Given the cost of an eigendecomposition, this is advantageous especially when performed for a coherent QPE, which requires not just one, but $b$ ancilla qubits, making the effort of simulation $\mathcal{O}(G2^{n+2b})$. In this case we have a scaling advantage for $b>n$.
 
Which of these approaches is more efficient depends on the required precision and the size of the matrix. An analysis of this trade-off and the respective timing results are presented in section \ref{sec:results}.

\subsection{Measurements}
Finally, emulators have an advantage over actual quantum computers when it comes to estimating the expectation values of measurements. On a quantum computer, a measurement of $n$ qubits only yields $n$ bits of information, returning one of the states $i$ as the result with probability given by $|\alpha_i|^2$. Our classical simulations are $\mathcal{O}(2^n)$ times more expensive than running the algorithm on a quantum computer, as we have to operate on the exponentially large vector representation. However, in return, we get the complete distribution of measurements and not just a single measurement sample.

While a quantum computer will often have to repeat an algorithm many times to get a (statistical) measurement with high enough accuracy, the classical emulation of such repeatedly executed measurements can easily be done in one step and the expectation value can immediately be evaluated. This removes the need for sampling and hence greatly reduces the overall simulation time.

As the time savings of emulation compared to simulation are just the number of repetitions of the circuit, no benchmark measurements are needed.

\section{Performance results}\label{sec:results}

\subsection{Experimental Setup}
\label{subsec:setup}

We compare performance quantum simulation and emulation on several systems. 

For distributed quantum FFT and phase estimation we use Stampede\cite{Stampede} system at the Texas Advanced Computing Center (TACC)/Univ. of Texas, USA ($\#10$ in the current TOP500 list). It consists of 6400 compute nodes, each of which is equipped with two sockets of Xeon E5-2680  connected via QPI and 32GB of DDR4 memory per node (16GB per socket), as well as one Intel\textregistered  Xeon Phi\texttrademark  SE10P co-processor. Each socket has 8 cores, with hyperthreading disabled. We use OpenMP~4.0~\cite{openmp2013openmp} to parallelize computation among threads. The nodes are connected via a Mellanox FDR 56 Gb/s InfiniBand interconnect. We have used  $\text{Intel}^{\textregistered}$ Compiler v15.0.2, $\text{Intel}^{\textregistered}$ Math Kernel Library (MKL) v11.2.2, and $\text{Intel}^{\textregistered}$  MPI Library v5.0.

Additional single-node and single-core benchmarks were done on an $\text{Intel}^{\textregistered}$ Core\texttrademark i7-5600U processor, unless specified otherwise. 

\subsection{Arithmetic Operations and Mathematical Functions}

\begin{figure}[t]
	\centering
	\resizebox{\linewidth}{!}{\input{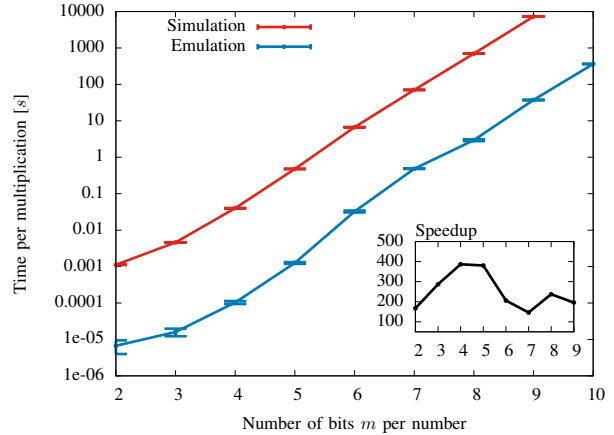}}
	\caption{Timings for emulation and simulation of a multiplication of two $m$-qubit numbers into a third register consisting of $m$ qubits (requiring a total of $n=3m$ qubits). There is a clear speedup when emulating this operation instead of simulating it at gate-level.}
	\label{fig:qmathmulttiming}
\end{figure}

All experiments for arithmetic operations were performed on a single core of an Intel Xeon E5-2697v2 processor due to the tremendous overhead in time when performing calculations with numbers consisting of more qubits than one node can handle. Such cases can only be dealt with by emulating the classical function, which effectively performs one global permutation of the (distributed) state vector.

Figure \ref{fig:qmathmulttiming} shows performance results comparing the runtimes of simulating and emulating a multiplication of two $m$-bit integers $a$ and $b$ into a third register $c$. The advantage of the emulator, performing more than one hundred times faster, can clearly be seen.

\begin{figure}[b]
	\centering
	\resizebox{\linewidth}{!}{\input{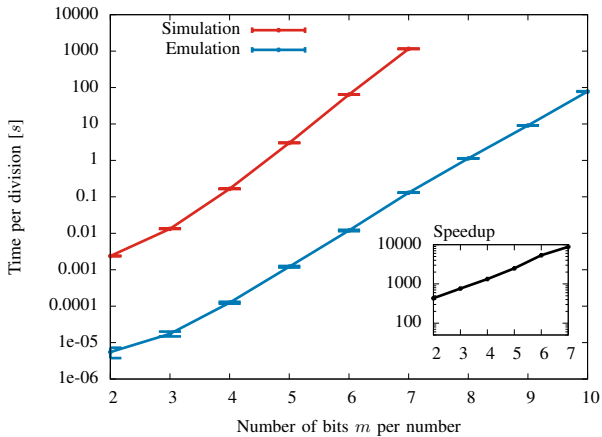}}
	\caption{Timings for emulation and simulation of an integer division of two $m$-qubit numbers into a third register consisting of $m$ qubits (requiring a total of $n=3m$ qubits). The speedup is far greater than for multiplication, which is due to the extra work qubits required to do the test for less/equal by checking for overflow. In addition, the numbers used for the division are limited to 7 bits due to the larger memory requirements caused by the extra work qubits.}
	\label{fig:qmathdivtiming}
\end{figure}

A much larger advantage can be seen for division, which requires additional work qubits to perform the calculation. This incurs an exponential cost on a simulator. In Figure \ref{fig:qmathdivtiming}, the runtime advantage for a division can be seen. It can be observed that the overhead grows with the number of qubits used to represent the integers, as the number of required work-qubits grows as well.

Even more dramatic effects can be expected when dealing with complex mathematical operations such as trigonometric functions, where some kind of series expansion or iterative procedure with many intermediate results is used. For each of these temporary values, additional $m$ qubits are required, causing an exponential overhead of the simulation in both space and time. Emulating such classical reversible functions not only pays off but makes it feasible on today's classical supercomputers, which otherwise would not be able to handle the enormous memory requirements.

\subsection{Quantum Fourier Transform}

\begin{figure}[t]
	\centering
	\resizebox{\linewidth}{!}{\input{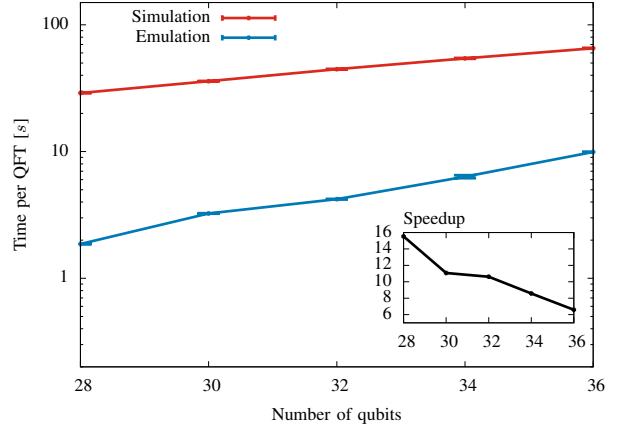}}
	\caption{Execution times for emulation and simulation of a quantum Fourier transform of $N$-qubits. Both emulation and simulation are run on $2^{N-28}$ nodes in order to keep the problem size per node constant. The emulator shows a clear advantage even when executing the QFT on a large number of qubits.}
	\label{fig:qfttiming}
\end{figure}

To benchmark the quantum Fourier transform, we use parallel implementations of both the simulator and the emulator, storing the wave function for $28$ qubits locally and using $2^{N-28}$ nodes for $N(\ge 28)$ qubits, which corresponds to weak scaling (keeping the problem size per node constant). In Figure~\ref{fig:qfttiming}  one can clearly see that simulating the QFT circuit is worse than directly performing a one-dimensional distributed classical fast Fourier transform. For the latter we used Intel Cluster FFT from Intel MKL library, which we found to be faster than FFTW~\cite{FFTW05}.

We observe that quantum emulation is $15\times$ faster than quantum simulation on a single node. Equations~\ref{eq:tqft} and~\ref{eq:tfft} show that FFT is expected to be faster than QFT on a single node by a factor of $\frac{n\text{FLOPS}_\text{achieved}}{\text{B}_\text{mem}}$, where $n$ is number of qubits (28 in our case). On a single node, FFT achieves $\approx20$ Gflops. Since $\text{B}_\text{mem}$ is 40 GB/s, the expected speedup is $28\cdot20/40=14$, which is very close to the observed speedup.

As we increase both the system size and the number of nodes, we observe degradation of the weak scaling times. This is expected due to the increasing amount of communication. We also observe a drop in the speedup of FFT over QFT. As Equations~\ref{eq:tqft} and~\ref{eq:tfft} show, the ratio of communication times between QFT and FFT is $\log_2(P) / 3$. Hence, for 2 and 4 nodes, we expect FFT to communicate more than QFT, resulting in some degradation in speedup and indeed, it drops to $11\times$. As $P$ increases, we expect the speedup of FFT over QFT to increase. However, we observe further slowdown, which we believe is due to the fact that the all-to-all communication phase of FFT becomes more limited by network congestion and thus does not follow the simplistic performance model of Equation~\ref{eq:tfft}. Nevertheless, we find a substantial $6-15\times$ speedup due to emulation.

\subsection{Quantum Phase Estimation}

\newcommand{\ra}[1]{\renewcommand{\arraystretch}{#1}}
\begin{table*}[!t]
	\ra{1.3}
	\begin{center}
		\begin{tabular}{@{}llllllll@{}}\toprule
			\textbf{\# qubits $\mathbf n$ acted on by $\mathbf U$} & \textbf 8		& \textbf 9		& \textbf{10}	& \textbf{11}		& \textbf{12}		& \textbf{13}	& \textbf{14}\\
			\textbf{Number of gates} $\mathbf G$& \textbf{29}       & \textbf{33}       & \textbf{37}       & \textbf{41}       & \textbf{45}       & \textbf{49}       & \textbf{53}\\ \midrule
			$T_{\text{apply $U$ with simulator}}$ [s]& $1.44\cdot 10^{-4}$ & $1.60\cdot 10^{-4}$ & $1.80\cdot 10^{-4}$ & $2.11\cdot 10^{-4}$ & $2.44\cdot 10^{-4}$ & $3.46\cdot 10^{-4}$ & $4.92\cdot 10^{-4}$ \\
			$T_{\text{construction of dense }U}$ [s]& $7.60\cdot 10^{-4}$ & $3.46\cdot 10^{-3}$ & $1.55\cdot 10^{-2}$ & $6.88\cdot 10^{-2}$& $3.02\cdot 10^{-1}$ & $1.32$ & $5.69$ \\ 
			$T_{\text{zgemm of dense}U}$ [s] & $8.39\cdot 10^{-4}$ & $6.71\cdot 10^{-3}$ & $5.37\cdot 10^{-2}$ & $4.29\cdot 10^{-1}$ & $3.44$ & $2.75\cdot 10^{1}$ & $2.20\cdot 10^{2}$ \\ 
			$T_{\text{zgeev of dense }U}$ [s] & $9.60\cdot 10^{-2}$ & $5.27\cdot 10^{-1}$ & $1.70$ & $6.72$ & $3.22\cdot 10^{1}$ & $1.80\cdot 10^{2}$ & $9.01\cdot 10^{2}$ \\ 
			\\ \textbf{Cross-over [\# bits of precision]}& & & & & & &\\ \midrule
			Repeated Squaring &  6  & 9  & 12 & 15 & 18 & 21 & 24\\
			Eigendecomposition & 10 & 12 & 14 & 15 & 18 & 19 & 21 \\

			\bottomrule
			\\
		\end{tabular}
		\caption{Timings of the various steps involved when simulating or emulating a quantum phase estimation. Our example is for the time evolution of a one-dimensional transverse field Ising model. The lower panel shows the cross-over precision in bits at which emulation using repeated squaring or eigendecomposition becomes advantageous over direct simulation.}
		\label{tbl:qpe_results}
	\end{center}
\end{table*}

We have used the Intel MKL implementations of complex matrix-matrix multiplication (\texttt{zgemm}) and a general eigensolver (\texttt{zgeev}) to perform repeated squaring and to determine the eigensystem, respectively. For the QPE timings, we focus on single-node performance, due to the fact that the ScalaPACK implementation of \texttt{zgeev} scales very poorly with number of nodes. There exist more scalable general eigensolvers, such as e.g. FEAST~\cite{Tang:2014:NHP:2663510.2663511}, but a detailed analysis of such solvers is beyond the scope of this paper.
Table~\ref{tbl:qpe_results} depicts the results for applying QPE to a unitary operator $U$ acting on different numbers of qubits $n\in\{8,...,14\}$. For each $n$, we determine the number of bits of precision corresponding to the cross-over point at which emulation becomes faster than quantum simulation.

For small $n$, the cross-over points for \texttt{zgemm} are very close to $n$, which is a clear discrepancy from the analytical model (see subsection \ref{subsec:qpe}), which is due to the fact that the model ignores constant  overheads, such as achievable bandwidth and flops. The ratio of these two overheads, while constant, has a much higher impact for small values of $n$, resulting in a digression from the model. As $n$ increases, this impact decreases and the cross-over values begin to approach our analytical prediction. Similar arguments hold for \texttt{zgeev}.

\subsection{Comparison against other Simulators}

In order to show that the obtained speedups result from emulation and do not originate from a suboptimal implementation of our simulator, we provide benchmarks comparing the performance of our simulator to other state-of-the-art simulators, namely qHiPSTER \cite{smelyanskiy2016qhipster} and LIQ$Ui\Ket{}$ \cite{wecker14}.  

\begin{figure}[t]
	\centering
	\resizebox{\linewidth}{!}{\input{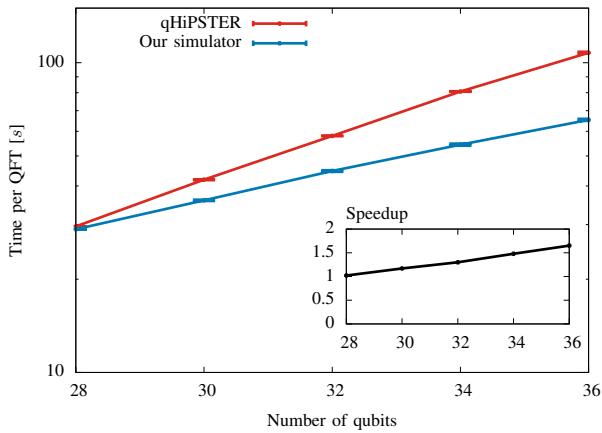}}
	\caption{Comparison between qHiPSTER and our simulator for applying  a quantum Fourier transform. The performance advantage of our simulator grows with the required communication, allowing simulations of larger systems.}
	\label{fig:qfttimingqhipster}
\end{figure}

The simulator benchmarks consist of two operations: Applying a QFT and an entangling operation, where the latter applies a Hadamard gate to the first qubit, followed by a series of CNOTs acting on all other qubits, all conditioned on the first qubit.

Since only qHiPSTER provides a distributed multi-node implementation, the parallel QFT comparison is exclusively carried out between qHiPSTER and our simulator. 
We show the weak-scaling behavior in Figure~\ref{fig:qfttimingqhipster}, where $N$ varies between 28 and 36 and the number of sockets is chosen to keep the memory per node constant (i.e. using from 1 to 256 nodes). Note that our parallel simulator shows a growing advantage as the requirement for communication increases. This stems from the fact that our simulator takes advantage of the structure of gate matrices, allowing e.g. to reduce the communication for diagonal gates such as the conditional phase shift.

\begin{figure}[ht]
	\centering
	\resizebox{\linewidth}{!}{\input{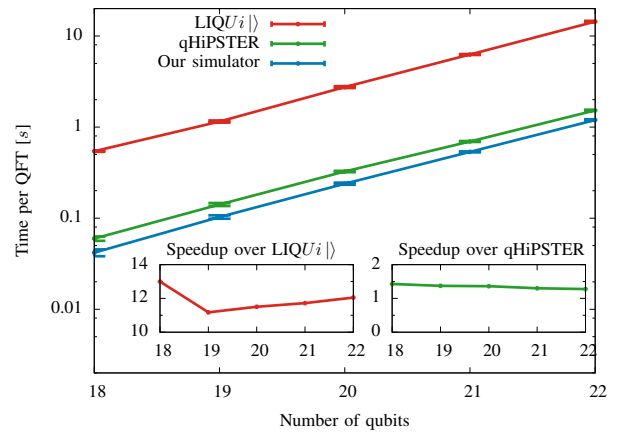}}
	\caption{Comparison of our simulator to qHiPSTER and LIQ$Ui\Ket{}$ for applying  a quantum Fourier transform on a single node. Our simulator clearly shows the best performance.}
	\label{fig:qfttimingliquid}
\end{figure}

\begin{figure}[ht]
	\centering
	\resizebox{\linewidth}{!}{\input{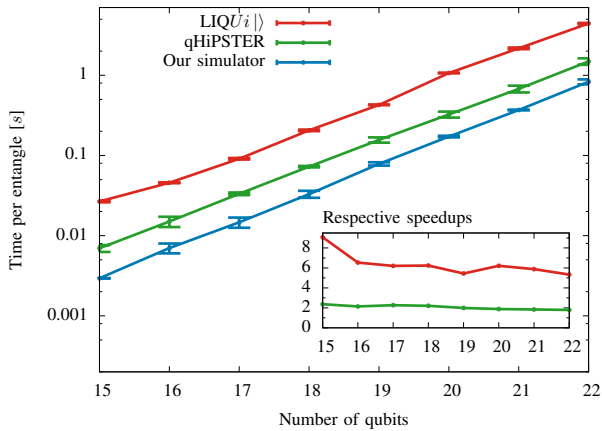}}
	\caption{Comparison of our simulator to qHiPSTER and LIQ$Ui\Ket{}$ for applying  an entangling operation on a single node. Our simulator achieves significant speedups of $2\times$ and $6\times$, respectively.}
	\label{fig:entangletimingliquid}
\end{figure}

The single node performance is depicted in Figure~\ref{fig:qfttimingliquid} for a QFT, and in Figure~\ref{fig:entangletimingliquid} for the entangling operation, which provides further proof of our simulator outperforming other ones. As a consequence, there will be an even larger advantage of our {\em emulator} against those simulators.

\section{Summary}\label{sec:summary}

The development of quantum algorithms that promise to solve important open computational problems has caused quantum computing to be viewed as a viable long-term candidate for post-exascale computing. Due to the current lack of universal quantum computers, the testing, debugging, and development of algorithms is done on classical systems, employing high-performance simulators. For the case of noiseless, perfect simulations, we propose to emulate the algorithms instead, making use of the performance optimizations presented in this paper. Yet, this emulation is only possible if the quantum program is available in a high-level language, where the higher levels of abstractions are easy to identify. This is the case in the compilation framework described in \cite{haener2016}, where emulators have been suggested at various levels, and can also be integrated into both the LIQ$Ui\Ket{}$  \cite{wecker14} and Quipper \cite{green13} quantum programming languages.

Our results show that emulating quantum programs allows to test and debug large quantum circuits at a cost that is substantially reduced when compared to the simulation approaches which have been taken so far. The advantage is already substantial for operations such as the quantum Fourier transforms, and grows to many orders of magnitude for arithmetic operations, since emulation avoids simulating ancilla qubits (needed for reversible arithmetic) at an exponential cost. Emulation will thus be a crucial tool for testing, debugging and evaluating the performance of quantum algorithms involving arithmetic operations, which includes quantum accelerated Monte Carlo sampling \cite{somma2008quantum} and machine learning applications \cite{wiebe2014quantum,wiebe2014quantumn,rebentrost2014quantum}.

\ifCLASSOPTIONcompsoc
  \section*{Acknowledgments}
\else
  \section*{Acknowledgment}
\fi

This work was supported by the Swiss National Science Foundation through the National Competence Center for Research NCCR QSIT and by Microsoft Research. Benchmark computations have been performed on Stampede at the Texas Advanced Computing Center (TACC)  and on the Euler cluster at ETH Zurich. The authors  thank G. Nadurra 
for inspiration.

\FloatBarrier



%
\bibliographystyle{IEEEtran}
\bibliography{references}

\end{document}